\documentclass[apj]{emulateapj}
\usepackage{amsmath}
\usepackage{mathrsfs}

\newcommand{\kms}{km\,s$^{-1}$}
\newcommand{\mearth}{$M_{\oplus}$}
\newcommand{\rearth}{$R_{\oplus}$}
\newcommand{\msun}{$M_{\odot}$}
\newcommand{\rsun}{$R_{\odot}$}

\newcommand{\ktt}{Kepler-32}
\newcommand{\eg}{{\it e.g.}}

\newcommand{\gcmc}{g\,cm$^{-3}$}
\newcommand{\gcms}{g\,cm$^{-2}$}

\newcommand{\kep}{{\em Kepler}}

\begin{document}
\slugcomment{Accepted for publication in the Astrophysical Journal}
\shorttitle{Characterizing the Cool KOIs IV: \ktt}
\shortauthors{Swift et al.}
\title{Characterizing the Cool KOIs IV: Kepler-32 as a prototype for
  the formation of compact planetary systems throughout the Galaxy} 
\author{Jonathan J. Swift\altaffilmark{*,1}, 
  John Asher Johnson\altaffilmark{1,2,3,4}, 
  Timothy D. Morton\altaffilmark{1,3}, 
  Justin R. Crepp\altaffilmark{5},
  Benjamin T. Montet\altaffilmark{1},
  Daniel C. Fabrycky\altaffilmark{6,7}, and
  Philip S. Muirhead\altaffilmark{1}}
\altaffiltext{*}{jswift@astro.caltech.edu}
\altaffiltext{1}{Department of Astrophysics, California Institute of
  Technology, MC 249-17, Pasadena, CA 91125} 
\altaffiltext{2}{Division of Geological and Planetary Sciences,
  California Institute of Technology, Pasadena, CA 91125} 
\altaffiltext{3}{NASA Exoplanet Science Institute (NExScI), CIT Mail
  Code 100-22, 770 South Wilson Avenue, Pasadena, CA 91125, USA} 
\altaffiltext{4}{Sloan Fellow}
\altaffiltext{5}{Department of Physics, 225 Nieuwland Science Hall,
  University of Notre Dame, Notre Dame, IN 46556} 
\altaffiltext{6}{Department of Astronomy and Astrophysics, University
  of California, Santa Cruz, Santa Cruz, CA 95064} 
\altaffiltext{7}{Hubble Fellow}

\begin{abstract}
  The \kep\ space telescope has opened new vistas in exoplanet
  discovery space by revealing populations of Earth-sized planets that
  provide a new context for understanding planet
  formation. Approximately 70\% of all stars in the Galaxy belong to
  the diminutive M dwarf class, several thousand of which lie within
  {\kep}'s field of view, and a large number of these targets show
  planet transit signals.  
  The \kep\ M dwarf sample has a characteristic mass of
  0.5\,\msun\ representing a stellar population twice as common as
  Sun-like stars. 
  \ktt\ is a typical star in this sample
  that presents us with a rare opportunity: five planets transit this
  star giving us an expansive view of its architecture.
  All five planets of this compact system orbit their host star within
  a distance one third the size of Mercury's orbit with the innermost
  planet positioned a mere 4.3 stellar radii from the stellar
  photosphere. 
  New observations limit possible false positive scenarios allowing us
  to validate the entire \ktt\ system making it the richest known
  system of transiting planets around an M dwarf.
  Based on considerations of the stellar dust sublimation radius, a
  minimum mass protoplanetary nebula, and the near period
  commensurability of three adjacent planets, we propose that the 
  \ktt\ planets formed at larger orbital radii and migrated inward to 
  their present locations. The volatile content 
  inferred for the \ktt\ planets and order of magnitude estimates for
  the disk migration rates suggest these planets may have formed
  beyond the 
  snow line and migrated in the presence of a gaseous disk. If true,
  this would place an upper limit on their formation time of $\sim
  10$\,Myr. 
  The \ktt\ planets are representative of the full ensemble of planet
  candidates orbiting the \kep\ M dwarfs for which we calculate an
  occurrence rate of $1.0 \pm 0.1$ planet per star. The
  formation 
  of the \ktt\ planets therefore offers a plausible blueprint for the 
  formation of one of the largest known populations of planets in our
  Galaxy. 
\end{abstract}

\keywords{ planetary systems --- methods: statistical --- planets and
  satellites: formation --- planets and satellites: detection ---  
   stars: individual (KID 9787239/KOI-952/Kepler-32)} 

\section{Background}
Before the discovery of exoplanets around main sequence stars two
decades ago, models of planet formation were based on a solitary
example: our own Solar System. Despite the discovery of hundreds, if
not thousands of additional planets in the years since then,
observational and theoretical efforts have focused on the formation of
planets around Sun-like stars. However, the Sun is not a typical 
star. Seventy percent of stars in the Galaxy are dwarfs of the M
spectral class \citep[``M dwarfs'' or ``red dwarfs,''][]{boc10}, with
masses that are only $\sim 10-50$\% the mass of the Sun, much cooler
temperatures, and different evolutionary histories. These differences
likely result in different formation and evolutionary histories for
their planets. For example, both Doppler and transit surveys have
revealed a paucity of gas giant planets around M dwarfs, and a
relative overabundance of planets with masses less than that of
Neptune \citep{how12}. This is in contrast to the high gas giant
occurrence rate around stars more massive than the Sun
\citep{john10a}. These correlations between stellar mass and gas giant
occurrence are likely a consequence of the lower disk masses around M
dwarfs, which result in less raw material available for planet
building \citep{lau04,ken08b}.

Roughly 5500 of the 160,000 stars targeted by NASA's \kep\ mission are 
M dwarfs with a mass distribution skewed toward the high mass end of
the spectral class (see \S\,\ref{ensemble} for details). Of these
stars, 66 show at least one periodic planetary transit signal. Aside
from a single outlier---the hot Jupiter system around 
KOI-254 \citep{joh12}---the ensemble of 100 planet candidates around
these M dwarfs have radii ranging from $\sim 0.5$ to 3\,\rearth\ and
semimajor axes within about a few tenths of an astronomical unit
\citep{mui12b}. These compact planetary configurations 
around the lowest mass stars \citep[\eg,][]{mui12a} offer a number of
advantages: the signal-to-noise ratio of the transit signals of
close-in planets is boosted due to the increased number of transits
per observing period \citep{gou03}; transit depths are larger for a
given planet radius allowing detections of ever smaller planets
\citep{nut08}; and the reduced temperatures and luminosities of their
host stars lead to equilibrium temperatures comparable to the Earth's,
despite their extreme proximity to their stars \citep{kas93,tar07}. 

The \kep\ M dwarf planets have been calculated to have an occurrence
rate a factor of $\sim 3$ higher than for solar-type stars with
occurrence rates increasing as planet size and stellar mass decrease 
\citep{man12,how12}. 
Microlensing studies also suggest a high planet occurrence around low
mass stars \citep{cas12}. However, the large errors on these results
and the fact that microlensing surveys probe planets at larger
separations from their host stars than transit surveys complicate the
comparison between samples.
The high frequency of small planets around low-mass stars is
compounded by the fact that lower mass stars are more common than
solar-type stars \citep{cha03}.  
Therefore the mechanisms by which small planets form around the lowest
mass stars determine the characteristics of the majority of planets
currently known to exist in our Galaxy.

\ktt\ is an M1V star with half the mass and radius of the Sun, roughly
two-thirds the Sun's temperature, and 5\% of its luminosity
\citep{mui12b}. While showing 5 distinct transit signals, to date only
two of the planets have been validated, \ktt\,b and c, from the timing
variations of their transit signals \citep[TTVs;][]{fab12a}. 
The \ktt\ host star and its planets are typical of the full
\kep\ M dwarf sample except for the chance alignment of this system
with respect to our line of sight that offers a unique and expansive
view of its dynamical architecture. To capitalize on this chance
alignment we bring to bear a suite of ground-based observations using
the W. M. Keck Observatory and the Robo-AO system on the
Palomar 60-inch telescope \citep{bar12} to validate and characterize
the system in detail. 
We follow this with an in-depth analysis that allows us to place tight
constraints on the formation and evolution of the \ktt\ planets, which
in turn has implications for the formation of planets around early M
dwarfs in general.

Since all planet parameters are derived directly from the physical
characteristics of the host star, we start by refining the stellar
properties in \S\,\ref{starprops}. High spatial resolution images and
optical spectra are then used by our false positive statistical
analyses to validate the remaining three transit signals from the
\kep\ data in \S\,\ref{validation}. These above analyses provide the
foundation for an accurate characterization of the \ktt\ planets
followed by an investigation into the formation and evolutionary
history of this system implied by its observed architecture in
\S\,\ref{planetsys}. We then turn to a discussion of the ensemble of
\kep\ M dwarf planets in \S\,\ref{discussion} and argue that the
formation pathway deduced for \ktt\ offers a plausible blueprint for
the formation of planets around the smallest stars, and thus the
majority of all stars throughout the Galaxy. 

\begin{center}
\begin{deluxetable}{lcl}
\tablewidth{2in}
\tablecaption{Observed Stellar Properties}
\tablehead{
\colhead{Parameter} &
\colhead{Value} &
\colhead{Ref.}}
\startdata
$\alpha$ (J2000)\dotfill    & 19:51:22.18  & 1 \\ 
$\delta$ (J2000)\dotfill    & $46^\circ 34^\prime 27^{\prime\prime}$ & 1 \\
$\mu_\alpha$ (mas/yr)\dotfill & -8  & 1 \\
$\mu_\delta$ (mas/yr)\dotfill & 20  & 1 \\
$K_P$\dotfill & 15.801 & 1\\
$A_V$\dotfill & 0.154  & 1 \\
$g$\dotfill & $17.251$ & 1 \\
$r$\dotfill & $15.913$ & 1 \\
$V$\dotfill & $16.452$\tablenotemark{*} & 2 \\
$J$\dotfill  & $13.616\pm0.023$  & 3 \\
$H$\dotfill   & $12.901\pm0.024$ & 3 \\
$K_s$\dotfill & $12.757\pm0.024$ & 3 \\
$v_{rad}$ (\kms) \dotfill & $-32.5 \pm 0.5$ & 4 \\
\enddata 
\tablerefs{(1) \cite{bor11b}; (2) \cite {jes05}; (3) \cite{bro11}; (4)
this work.}
\tablenotetext{*}{Converted from $g$ and $r$}
\label{star_tab_obs}
\end{deluxetable}
\end{center}

\section{\ktt\ Stellar Properties}\label{starprops}
The observed stellar properties for \ktt\ are summarized in
Table~\ref{star_tab_obs}. It was originally listed in the Kepler Input
Catalog (KIC) with an effective temperature of 3911K, below the
4500\,K threshold where the photometric method for determining stellar
parameters is deemed reliable \citep{bro11}. Follow-up,
medium-resolution infrared spectroscopy presented by \cite{mui12b}
revised and refined the KIC values to $T_{\rm eff} = 3727^{+102}_{-58}$
and [Fe/H]$ = 0.04 \pm 0.13$. 

In this work we supplement these measurements with an independent
estimation of the stellar mass, radius and metallicity using the
broadband photometric method presented by \cite{joh12}.
This method evaluates the posterior probability distribution of each
stellar parameter conditioned on the observed apparent magnitudes and
colors of the star listed in the KIC using model relationships such as
the \cite{del00} mass-luminosity relation and the \cite{wes05}
color-spectral-type relation. Our Markov Chain Monte Carlo analysis
yields $M_\star = 0.57 \pm 0.06$\,\msun, $R_\star = 0.53 \pm
0.04$\,\rsun, and [Fe/H]$ = -0.05 \pm 0.17$.

We combine the results of the above analyses with those of
\cite{bor11a}, \cite{mui12b}, and \cite{fab12a} by weighted mean to
obtain the final values used in this work presented in
Table~\ref{star_tab_prop}. We adopt 10\% errors on the \cite{bor11a}
values for stellar mass and radius. For the effective temperature, we
construct probability distribution functions and compute the mean and
$1\,\sigma$ errors numerically to account for the asymmetric errors in
the \cite{mui12b} value. Using the \cite{del00} $K$-band
mass-luminosity relationship we find that our final stellar mass and
its uncertainty give a distance modulus of $7.407 \pm 0.098$,
corresponding to a distance of $d = 303 \pm 14$\,pc.

\subsection{Stellar Age}
It is difficult to measure the age of field M dwarfs unassociated with
known moving groups or clusters because their global stellar
properties change little over the course of their main sequence
lifetime. However, \ktt\ has a slow rotation rate of $P = 37.8 \pm
1.2$\,days measured from the modulation of the light curve by star
spots \citep{fab12a} implying that it is old. 
To estimate the age of \ktt\ we use the age equation from
\cite{bar10}, 
\begin{equation}
t = \frac{\tau}{k_C}\ln\left(\frac{P}{P_0}\right) + 
\frac{k_I}{2\tau}\left(P^2-P_0^2\right).
\label{age_eq}
\end{equation}
The dimensionless constants $k_C = 0.646$\,days\,Myr$^{-1}$ and $k_I =
452$\,Myr\,day$^{-1}$ are determined observationally, the convection
turnover time is estimated as $\tau = k_I/2 \cdot f^2(B-V)$, where we
use $f(B-V) = 0.77(B-V-0.4)^{0.6}$ \citep{bar07} and $(B-V) = 1.5$
\citep[converted from $g-r$ using][]{jes05} to give $\tau =
155$\,days. Using the median value of the initial spin period, $P_0 =
2.81$, as required to produce the observed rotation rates for $\approx
0.6$\,\msun\ stars in the Praesepe cluster \citep{agu11} we derive an
age of 2.7\,Gyr for \ktt. The large scatter in observed rotation rates
for Praesepe members used here to calibrate $P_0$ suggests the age of
\ktt\ could be as young as 2.3\,Gyr or as old as 3.7\,Gyr. For this
work we will assume that the age of \ktt\ is greater than $2$\,Gyr.
\begin{center}
\begin{deluxetable*}{lcccccc}[!hb]
\tablewidth{0in}
\tablecaption{Derived Stellar Properties}
\tablehead{
\colhead{Parameter} &
\colhead{KIC\tablenotemark{1}} &
\colhead{M12\tablenotemark{2}} &
\colhead{F12\tablenotemark{3}} &
\colhead{this work} &
\colhead{Adopted Values}}
\startdata
$T_{\rm eff}$\,(K)       & $3911 \pm 200$   & $3727^{+102}_{-58}$    & $3900 \pm 200$   & \nodata            & $3793^{+80}_{-74}$  \\
$M_\star$\,(\msun)   & $0.49 \pm 0.05$  & $0.52 \pm 0.04$      & $0.58 \pm 0.05$   & $0.57 \pm 0.06$    & $0.54 \pm 0.02$ \\ 
$R_\star$\,(\msun)   & $0.56 \pm 0.06$  & $0.50 \pm 0.04$      &  $0.53 \pm 0.04$ & $0.53 \pm 0.04$    & $0.53 \pm 0.02$ \\
$[$Fe/H$]$          & $-0.056 \pm 0.2$ & $0.04 \pm 0.13$       & $0 \pm 0.4$      & $-0.05 \pm 0.17$   & $-0.01 \pm 0.09$ \\
$d$\,(pc)           & \nodata & \nodata & \nodata & $303 \pm 14$ & $303 \pm 14$ \\
age (Gyr)       & \nodata & \nodata & \nodata & $\sim 2.6$ & $> 2$ \\
\enddata 
\tablerefs{(1) \cite{bro11}; (2) \cite{mui12b}; (3) \cite{fab12a}} 
\label{star_tab_prop}
\end{deluxetable*}
\end{center}

\section{Validation of Transit Signals}\label{validation}
Two of the \ktt\ planets (b and c) have been validated previous to 
this study using signatures of their mutual dynamical interactions
\citep{fab12a}. The other three transit signals have hitherto retained
the status of planet candidates. The probabilities for \kep\ planet
candidates to be astrophysical false positives, such as blended
stellar eclipsing binaries, are generally low \citep{mor11}, and may
even be negligible in multiply transiting systems
\citep{lis12}. Diffraction limited images in the \kep\ band made with
the Robo-AO system on the Palomar 60-inch telescope
\citep{bar12} show no evidence of blended companions 
for a $5\,\sigma$ contrast of $\Delta z
\approx 3.5$ at $0^{\prime\prime}\!\!.5$ and $\Delta z
\approx 4.5$ at $1^{\prime\prime}$. High-resolution
optical spectroscopy with the HIRES spectrograph on 
Keck I and high-resolution adaptive optics imaging in the near
infrared using the NIRC2 camera on Keck II place stringent constrains
on the probabilities of various false positive scenarios, allowing
a probabilistic validation of the transit signals using the procedure
of \cite{mor12}. 

\subsection{Keck I/HIRES Spectroscopy}\label{KeckI}
We obtained spectra of \ktt\ at Keck Observatory using the
HIgh-Resolution Echelle Spectrometer \citep[HIRES,][]{vog94} on Keck I
with the standard observing setup used by the California Planet Survey 
\citep{john10b} which covers a wavelength range from 3640\AA\ to
7820\AA. Because of the star's faint visual magnitude we used the C2
decker corresponding to a projected size of $14^{\prime\prime}\!\!.0
\times 0^{\prime\prime}\!\!.851$ to allow sky 
subtraction and a resolving power of $R = 55000$. We obtained 2
observations of \ktt\ (UT 18 June 2011 and UT 13 March 2012), both
with an exposure time of 700\,seconds, each resulting in a
signal-to-noise ratio (SNR) of 20 at 6500\AA.  

To measure the systemic radial velocity of \ktt\ and to check for
evidence of a second set of stellar lines in the spectrum, we
performed a cross-correlation analysis with respect to a HIRES
spectrum of a similar star \citep[HIP 86961: $V_{rad} = -28.9 \pm
0.4$\,\kms;][]{nid02} using the methodology described by
\cite{john10a}. Our analysis yields a systemic radial velocity of
$-32.5 \pm 0.5$\,\kms. No second set of lines are evident above the
noise floor in the cross-correlation function, allowing us to rule out
stellar companions with $V > 10$\,\kms\ and $V$-band brightnesses
within 2 magnitudes.

\subsection{Keck II/NIRC2 Adaptive Optics Imaging}\label{KeckII}
We performed adaptive optics imaging using the NIRC2 instrument on the
Keck II telescope on UT 24 June 2011 to rule out false positive
scenarios involving blended sources. With a \kep\ bandpass magnitude
$K_P = 15.80$, \ktt\ is relatively faint for natural guide star
observations. Nevertheless, we were able to close the adaptive optics
(AO) system control loops on the star with a frame rate of
41\,Hz. With sufficient counts in each wavefront sensor subaperture, a
stable lock was maintained for the duration of the observations.
\begin{figure}[!hb]
\centering
\includegraphics[angle=90,width=3.4in]{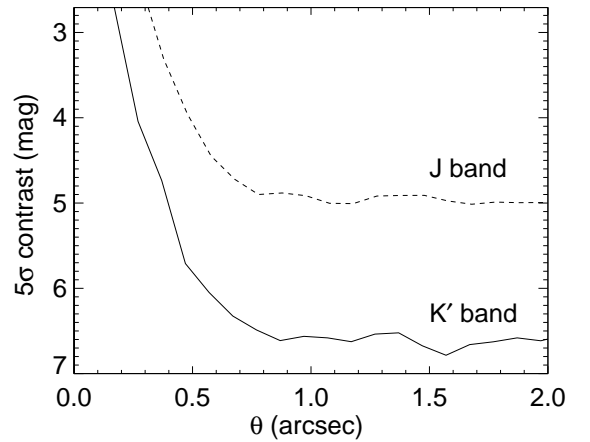}
\caption{$J$ and $K^{\prime}$ band contrast curves derived from our
  Keck II/NIRC2 adaptive optics imaging constrain false positive
  scenarios involving blended sources.} 
\label{contrast_curve}
\end{figure}

We acquired two sets of dithered images using the NIRC2 medium camera
(plate scale = 20\,mas\,pix$^{-1}$) a sequence of 9 images in the
$J$-filter (central wavelength = 1.25\,$\mu$m), and a sequence of 9
images in the 
$K^\prime$-filter (central wavelength = 2.12\,$\mu$m). Each image had
an exposure time of 20\,s, resulting in a total on-source integration
time of 180\,s per filter. To process the data, hot pixels were
removed, the sky-background was subtracted, and the images were
aligned then coadded. 
We measured the contrast achieved by NIRC2
adaptive optics imaging by comparing the peak intensity of the
star in the final processed image to the intensity of residual
scattered starlight at small angular separations. Specifically, we 
calculate the standard deviation of flux, $\sigma$, within a box of 
size $3 \times 3$\,FWHM, where FWHM is the point-spread function
full-width at half-maximum spatial scale (also the size of a speckle). The 
standard deviation is evaluated at numerous locations close to the
star and the results are azimuthally averaged to estimate the local
radial contrast profile. Figure~\ref{contrast_curve} shows the
$5\,\sigma$ contrast curves based on each reduced AO image. The full
instrument field of view is
$20^{\prime\prime}\times20^{\prime\prime}$, which corresponds to 5
\kep\ pixels on a side. No obvious contaminants were identified within
$\Delta K^\prime = 6.5$ at a separation of $0^{\prime\prime}\!\!.7$ or
farther from \ktt.

\subsection{\kep\ Light Curves}
Our light curve data come from the \kep\ space telescope, which is
conducting a continuous photometric monitoring campaign of a target
field near the constellations Cygnus and Lyra. A 0.95-m aperture
Schmidt telescope feeds a mosaic CCD photometer with a $10^\circ
\times 10^\circ$ field of view \citep{koc10,bor11b}. Data reduction
and analysis is described by \cite{jen10a,jen10b} and photometric and
astrometric data were made publicly available as part of the 28 July
2012 public data release. We downloaded the data from the Multimission
Archive at STScI (MAST), and we use the pipeline-corrected light
curves from Quarters 0 through 9 \citep{bat12}.  

\begin{figure}
\centering
\includegraphics[angle=0,width=3.4in]{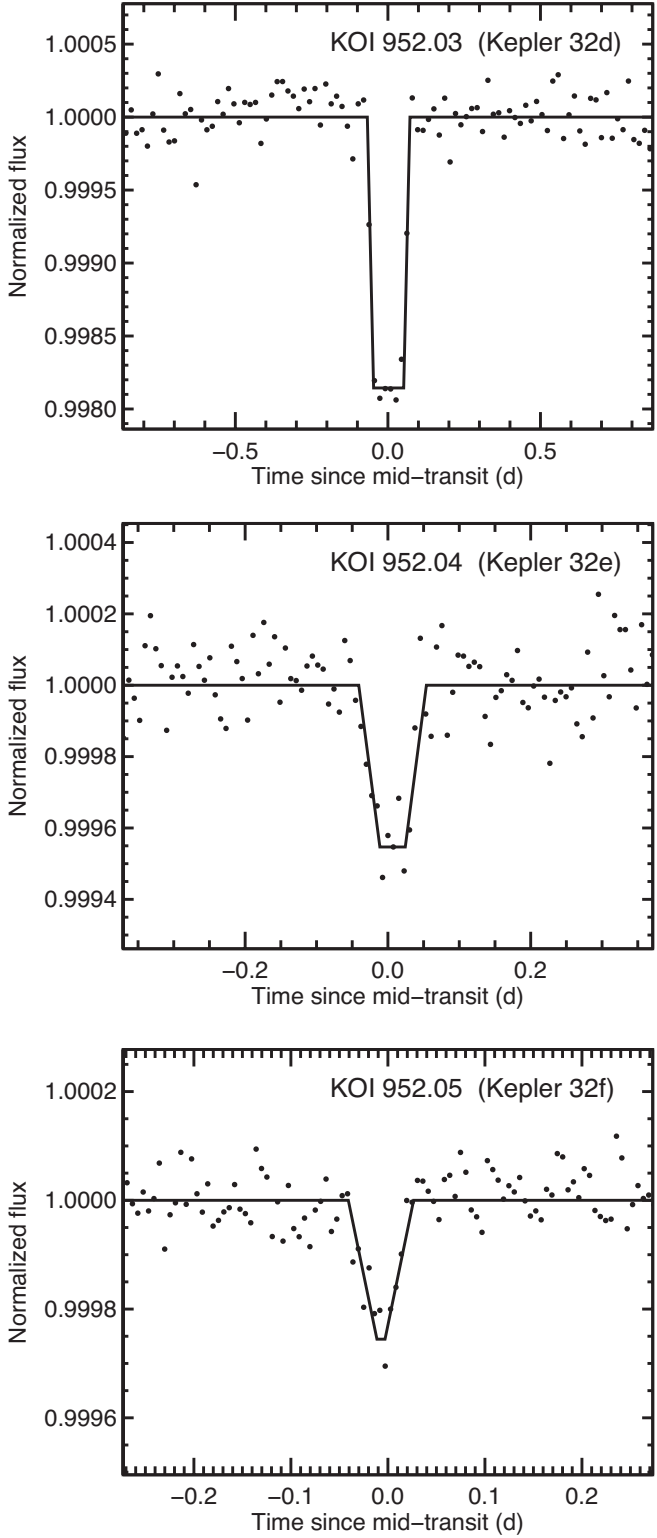}
\caption{Detrended, folded and binned \kep\ lightcurves for the three
  planets that we validate with our FPP analysis. The trapezoidal fits
  from the analysis of \cite{mor12} are overlaid.}
\label{lightcurves}
\end{figure}

These light curves were aggressively detrended to take out all
astrophysical variation by first masking out all transit signals and
then sequentially fitting a low-order polynomial to the time series
data in 2 day chunks. These detrended data were then folded for each
\ktt\,d, e and f with the other planet signals masked according to the
ephemeris of \cite{fab12a}. These folded transit signals were used
for the FPP analysis. The binned photometric data are shown in
Figure~\ref{lightcurves} with the trapezoidal fits used to assess the
transit shape overlaid. 

\subsection{False Positive Probability Analysis}\label{FPP}
To calculate the relative probabilities of astrophysical false
positive and {\em bona fide} transiting planet scenarios, we follow the
methodology of \cite{mor12}, which compares the observed transit shapes
to those of simulated false positive populations, subject to the
available observational constraints. Although occurrence rates may be
as high as 10\% for planets in the radius bins corresponding to
KOI-952.03, .04, and .05 \citep{how12} we assume a very conservative
occurrence rate of 1\%. 
 
\begin{figure}
\centering
\includegraphics[angle=0,width=3.4in]{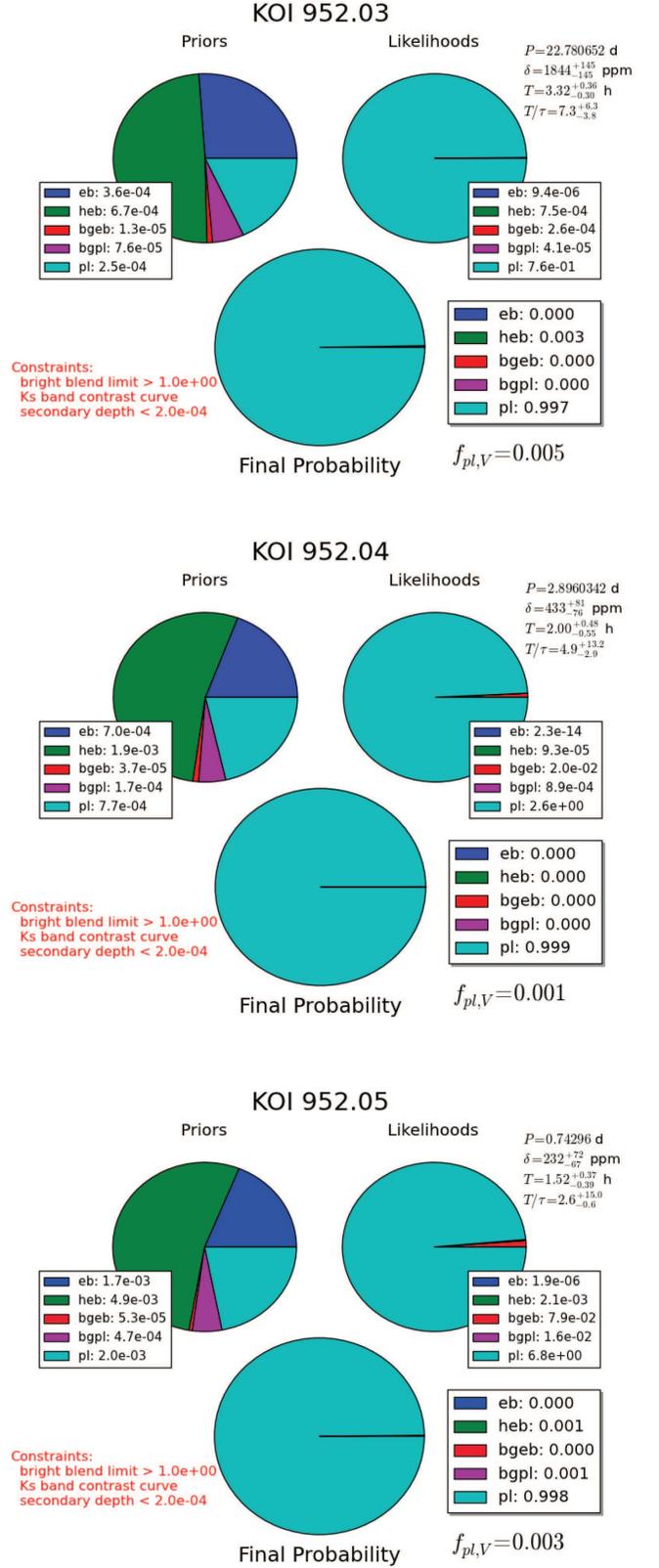}
\caption{The relative strength of each possible scenario in our
false positive calculation---eclipsing binary (eb), hierarchical
binary (heb), background eclipsing binary (bgeb), background planet
(bpl) and planet (pl)---is shown pictorially for our priors (upper left
chart), likelihood (upper right chart), and final probability (lower
chart) for each KOI-952.03 (top panel), .04 (middle panel), and .05
(bottom panel). Observational constraints used in the calculation are
stated in red (lower left) and the trapezoidal fit paramters are shown
(upper right) in each panel. The quantity $f_{pl,V}$ is the smallest
value of the planet occurrence that would result in validation for the
given transit signal.} 
\label{FPP2}
\end{figure}

Figure~\ref{FPP2} summarizes these FPP results for all three
signals. Despite our conservative estimates of planet occurrence rate
and without accounting for the fact that \ktt\,b and c
have already been confirmed to be coplanar transiting planets, which
would further decrease the FPPs by a factor of about 10, we estimate 
the FFPs for KOI-952.03, .04, and .05 to all be less than 0.3\%. We
therefore consider all five photometric signals to be fully validated
planet transits making \ktt\ the richest known system of
transiting planets around an M dwarf, and we assign the names
\ktt\,d, e, and f to the former candidates KOI-952.03, .04, and
.05, respectively. Verification of the transit signals allows us to
take full advantage of the fortuitous alignment of the \ktt\ system with
respect to our line of sight, and in the next section we look deeper
into the details and implications of its specific configuration.

\begin{center}
\begin{deluxetable}{lcccccc}
\tablewidth{0in}
\tablecaption{Derived Planet Properties}
\tablehead{
\colhead{KOI} &
\colhead{{\it Kepler}} &
\colhead{$P$\tablenotemark{1}} &
\colhead{$a$} &
\colhead{$R_p$} &
\colhead{$T_{eq}$\tablenotemark{a}} \\
\colhead{} &
\colhead{} &
\colhead{($days$)} &
\colhead{($AU$)} &
\colhead{($R_\oplus$)} &
\colhead{($K$)} }
\startdata
 952.05 & 32f  & 0.74296(7) & 0.0130(2)  & 0.81(5) & 1100 \\ 
 952.04 & 32e  & 2.8960(3)  & 0.0323(5)  & 1.5(1)  & 680 \\ 
 952.01 & 32b  & 5.9012(1)  & 0.0519(8)  & 2.2(1)  & 530 \\ 
 952.02 & 32c  & 8.7522(3)  & 0.067(1)   & 2.0(2)  & 470 \\ 
 952.03 & 32d  & 22.7802(5) & 0.128(2)   & 2.7(1)  & 340 \\ 
\enddata 
\tablenotetext{a}{assuming a Bond albedo $\alpha = 0.3$}
\tablerefs{(1) \cite{fab12a}}
\label{plan_tab}
\end{deluxetable}
\end{center}

\section{The \ktt\ Planetary System}\label{planetsys}
\begin{figure*}
\centering
\includegraphics[angle=0,width=5in]{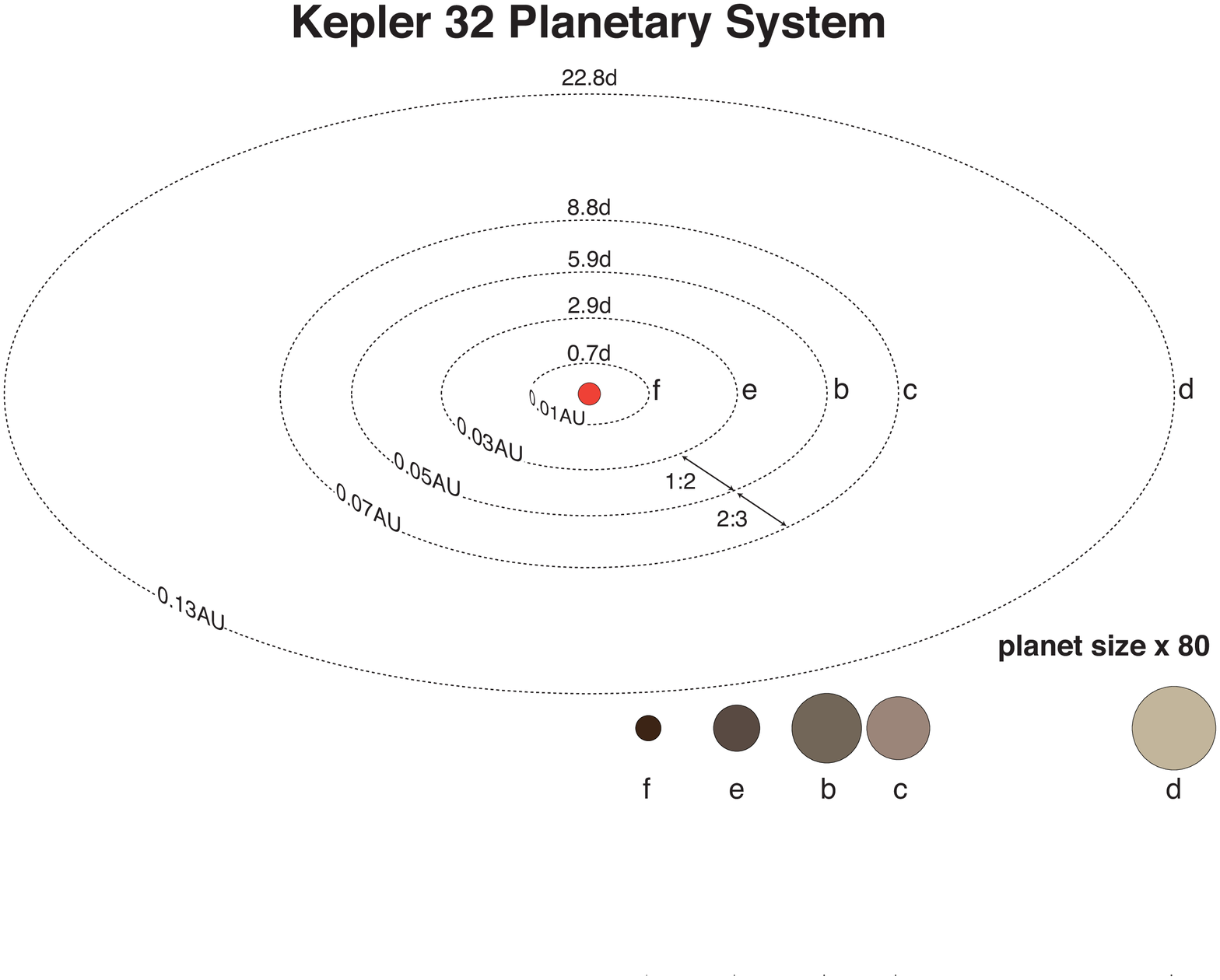}
\caption{Depiction of the Kepler-32 planetary system with the star and
  orbits drawn to scale. The relative sizes of the planets are shown
  at the bottom of the figure scaled up by a factor of $80$ in
  relation to their orbits.}
\label{kepler_32}
\end{figure*}

The increased accuracy with which we derive the stellar parameters for
\ktt\ (see Table~\ref{star_tab_prop}) leads to more precise planetary
parameters, which we outline in Table~\ref{plan_tab}.
The planets of \ktt\ have radii that are similar to values found in
the Solar System, from below Earth size up to about 70\% of
Neptune. However, the planetary system of \ktt\ has a remarkably
distinct dynamical architecture in comparison. Figure~\ref{kepler_32}
shows the relative sizes of the planets and their orbits, along with
labels denoting their periods, semimajor axes and period
commensurabilities rounded to the nearest integer. The five planets of
this system orbit within 0.13\,AU from the star, or approximately one
third of Mercury's
semimajor axis. The outermost planet lies within a region where the
stellar insolation is similar to Earth's, while at the other extreme,
the innermost planet orbits only 4.3 stellar radii from the
photosphere of \ktt. The three middle planets---termed e, b,
and c in order of increasing semimajor axis---exhibit period ratios
within 2\% of a 1:2:3 commensurability. 

\subsection{Tidal Evolution}

The proximity of the \ktt\ planets to their host star imply that their
dynamics have been significantly altered due to tidal forces induced
by the host star over the age of the system. Using an initial spin
period of 10 hours, a tidal dissipation factor of 100, a rigidity
factor of $3 \times 10^{11}$ and a density profile similar to Earth
such that the moment of inertia $I \approx 0.5\,M_pR_p^2$, we derive
the tidal locking timescales for the \ktt\ planets to all be $\lesssim
1$\,Myr \citep{mur00}. This timescale depends linearly on the
dissipation factor, $Q$, which is highly uncertain for
exoplanets. However, for a reasonable range of $Q$ values ranging up
to $10^5$ it can be expected that the rotation periods of all the
planets in the \ktt\ system are equal to their orbital period barring
spin-orbit resonances. 


Tidal forces from the host star will also damp the eccentricities of
the planets in direct proportion to $Q$ \citep{mur00}. For
$Q < 10^4$ the inner planets will have no free (primordial)
eccentricity. Though even for a low $Q$, the tidal damping time for
\ktt\,d, the most distant planet, is approximately 4\,Gyr, comparable
to the age of the system. The free eccentricities of \ktt\,b and c
have recently been estimated to have non-zero values from the phase of
the TTV signals \citep{lit12,wu12}. This suggests high $Q$ values for
these planets, consistent with their moderate inferred densities (see
\S\,\ref{TTV}).

Contrary to the trend for \kep\ pairs to lie longward of mean
motion resonance \citep{fab12b}, \ktt\,b and c lie 1.1\% 
{\it shortward} of resonance. The high $Q$ values suggested above
would limit dissipative mechanisms that may spread their orbits
\citep{lit12a,baty12}. Also \ktt\,b is near a 2:1 resonance with
\ktt\,e. However, our n-body simulation of \S\,\ref{architecture}
suggests planet e has too small an effect to account for these effects.

\subsection{Mass and Density Estimates} \label{TTV}
As mentioned above, \ktt\,b and c were validated based on the TTVs
observed due to their mutual interactions \citep{fab12a}. The
amplitude of the TTV signals specify upper limits to their masses of
be 6.6 and 8.4\,\mearth, implying densities of less than 3.4 and
5.7\,\gcmc\ for b and c respectively \citep{lit12}. However, the TTV
phases are measured to be $\sim 45^\circ$ from $0$ and $\pi$ implying
a small free eccentricity. A correction for this effect by
\cite{wu12} reduce the estimated masses for \ktt\,b and c by factors
of $0.51$ and $0.45$. With these corrections the masses and densities for
\ktt\,b and c are 3.4 and 3.8\,\mearth\ and 1.7 and 2.6\,\gcmc. The
nominal masses for \ktt\,b and c imply mass-radius relationships of $M
\propto R^{\gamma_p}$, with $\gamma_p = 1.5-1.9$, similar to the value
of 2.06 for the six Solar System planets bounded by Mars and Saturn
\citep{lis11}. 

The above stated densities imply that \ktt\,b and c are composed of a
significant amount of volatiles. Using Equations 7 and 8 of
\cite{for07}, we find that if \ktt\,b and c had no atmospheres they
would be expected to contain $\sim 96\%$ and $\sim 56\%$
volatiles. However, given the equilibrium temperatures of \ktt\,b and
c a large fraction of their volatile content likely exists in the form
of an atmosphere.

\subsection{Atmospheric Evolution} 
The proximity of the \ktt\ planets to their host star suggest
significant atmospheric evolution due to evaporation, outgassing or
both processes. 
The equilibrium temperature of \ktt\,f is $\sim
1100$\,K and its radius is measured to be 0.81\,\rearth. For a planet
this small with such a high equilibrium temperature the atmospheric
mass fraction would have to be very small, $\sim 10^{-5}$
\citep{rog11}. Using an extreme ultraviolet luminosity of
\ktt, $L_{EUV} \approx 10^{26.6}$ \citep{hod94} and following
\cite{lec07} using a conservative mass 
loss efficiency of $\epsilon_{UV} = 0.1$ we derive an atmospheric mass
loss of $\sim 10^8$\,g\,s$^{-1}$. Thus the timescale to lose its
atmosphere is more than 100 times shorter than the age of the
\ktt\ system. We therefore conclude the \ktt\,f contains no
atmosphere. 

Given the size and equilibrium temperature of \ktt\,e its atmospheric
mass fraction must also be small, $M_a/M_p \sim 10^{-4}$
while the present day atmospheric mass loss rate is between $10^7$ and 
$10^8$\,g\,s$^{-1}$. The timescale for the complete loss of the
\ktt\,e atmosphere is calculated to be between 0.2 and
2\,Gyr. Therefore \ktt\,e must have lost a significant fraction of any
atmosphere it started with.

The total atmospheric mass loss for the other three planets is at
least $\sim 10^{-4}$\,\mearth\ for reasonable choices of planetary
mass. If these planets have relatively low-density cores (ice and
rock) and started out with large atmospheres, they could have suffered 
considerable atmospheric evolution due to the heating by \ktt. Thus
the observed sizes of the \ktt\ planets are likely determined in part
by the extreme ultraviolet and X-ray luminosity of their host
star. However, the mass estimates from \S\,\ref{TTV} suggest that
\ktt\,b is 10\% less massive than \ktt\,c while being 10\% larger and
25\% closer to \ktt\ hinting that the mass-radius relation for the
\ktt\ planets is not determined solely by a simple atmospheric
evolution model.

\subsection{\ktt\ Planetary System Architecture}\label{architecture}
The physical characteristics of the \ktt\ planets are summarized in
Table~\ref{plan_tab} and the remarkably compact and orderly
architecture of the system is shown schematically in
Figure~\ref{kepler_32}. As mentioned above, three of the planets lie
within 2\% of a 1:2:3 period commensurability. \ktt\,e and b have a
period ratio of 2.038, which is 1.9\% longward of commensurability,
while \ktt\,b and c have a period ratio of 1.483, or 1.1\% shortward
of commensurability.

Planets within a mean motion resonance can stray a few percent from
commensurability and maintain the libration of resonant angles
\citep{mur00}. However without detailed knowledge of the individual
orbits it is not possible to determine with certainty if a planet
pair is in a resonant configuration. Therefore we assess the
significance of the near commensurability of \ktt\,e, b, and c
using a probabilistic argument.   

We randomly populate 5 planet systems with periods between the
inner and outermost planets in the \ktt\ system enforcing
separations larger than $2\sqrt3$ for every
pair of neighboring planets \citep{gla93} and larger than 9
for chains of planets \citep{cha96,smi09,lis11} in units of mutual
Hill radii. In this section a mass-radius relationship of  
$M \propto R^{2.06}$ \citep{lis11} is adopted for consistency. The final
ensemble of systems provide a baseline against which we can gauge the
significance of the \ktt\ architecture. The occurrence of 
period ratios in our randomly drawn sample that lie within either the
observed period ratio of \ktt\,e and b, or b and c, is fairly high,
11\% and 12\%, respectively. However the fraction of systems
arranged in a chain involving any combination of 1:2 or 2:3
near-commensurabilities is $1.9\% \pm 0.5\%$. This suggests that
the architecture we observe today reflects the result of dynamical
evolution rather than a result of chance. 

Given the compactness of the \ktt\ system, we expand on the dynamical
stability simulations of \cite{fab12a} by considering all five
planets. \ktt\,f is very likely on a circular orbit and lies $>30$  
mutual Hill radii away from the nearest planet. It therefore is not
expected to have any bearing on the stability of the system
\citep{smi09a} and is not included in the following analysis.

\begin{figure}
\centering
\includegraphics[angle=0,width=3.4in]{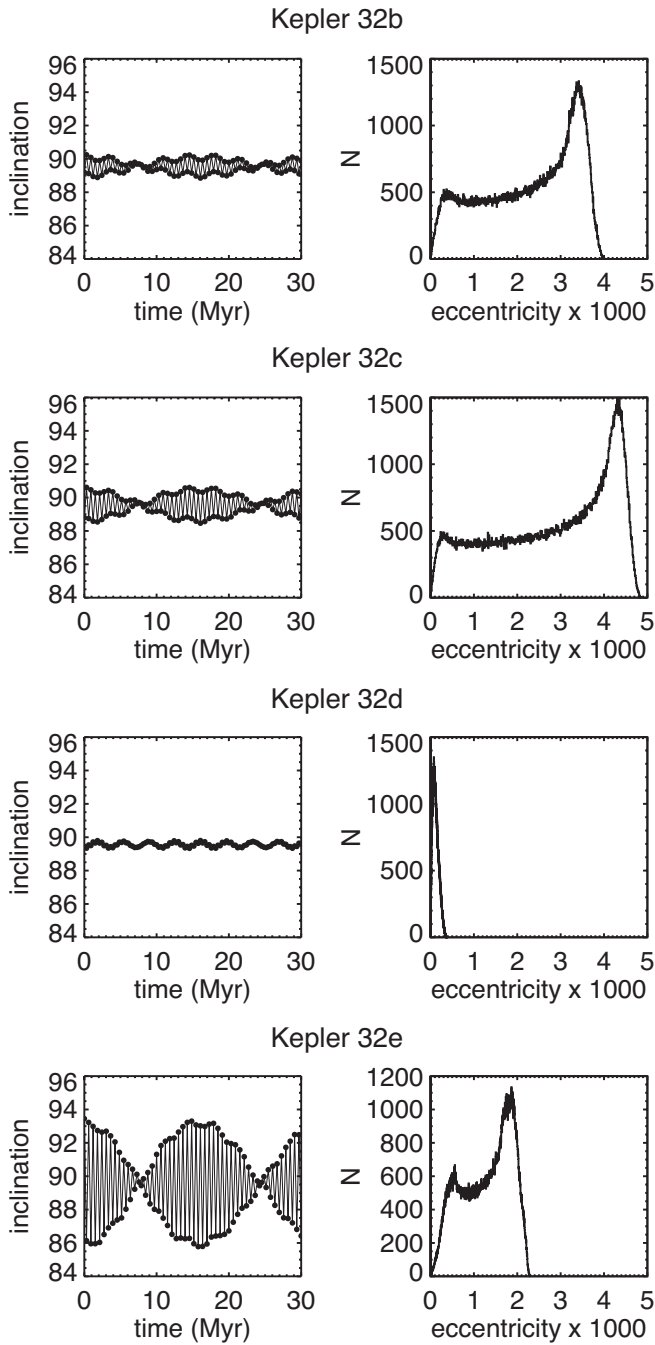}
\caption{Time evolution of the inclinations sampled every 100kyr and
  the distributions of eccentricities for the \ktt\ planets sampled
  every 100\,yrs from our n-body simulation. \ktt\,f was not included
  in the simulation due to its dynamical isolation from the rest of
  the planetary system (see text).}
\label{stability}
\end{figure}

Using the Mercury6 software \citep{cha99} employing the hybrid
symplectic Bulirsch-Stoer integrator we integrate the \ktt\ system for
30\,Myr, or $\sim 500$ million orbits of the outermost planet. Our
initial conditions were set according to the ephemeris of
\cite{fab12a} with an initial time step of one tenth the orbital
period of the innermost planet (in this case, \ktt\,e). We began with
zero eccentricity and random inclinations drawn uniformly between the
minimum and maximum values that would produce a transit from our
vantage point. We did not include any tidal effects. The
eccentricities and inclinations varied stably with small
amplitudes. Figure~\ref{stability} shows the inclination evolution
resampled at 100\,kyr time steps and histograms of the eccentricities
measured every 100\,yrs over the 30\,Myr simulation for the outer four
planets.

\subsection{The Formation and Evolution of the \ktt\ Planetary System}
\begin{figure*}[!ht]
\centering
\includegraphics[angle=0,width=6in]{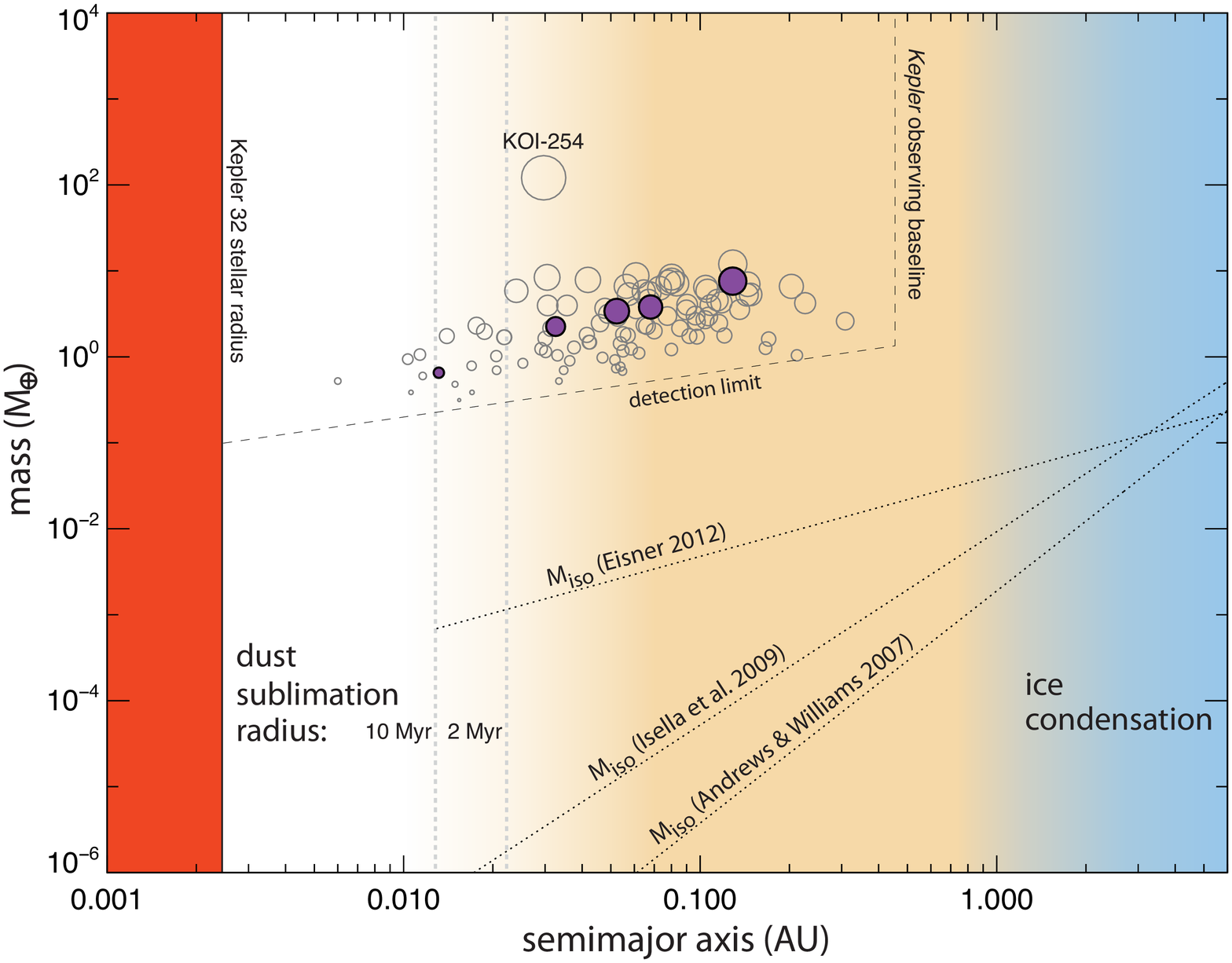}
\caption{The masses of the \ktt\ planets ({\em purple dots}) at their 
  respective semimajor axes are shown in relation to the full
  distribution of M dwarf planet candidate masses ({\em gray
    circles}) calculated according to $M \propto R^{2.06}$ except for
  \ktt\,b and c for which we use the masses from \S\,\ref{TTV}. The
  isolation masses of three different disk models are labeled 
  ($M_{iso}$, {\em dotted lines}). Also shown are conservative locations of
  the dust sublimation radius for pre-main sequence ages of 2\,Myr and
  10\,Myr \citep{bar98}, the approximate location of the ice line
  during the planet forming epoch \citep[{\em blue shaded
      region},][]{ken08b}, and the stellar radius of \ktt. The
  discovery space of the \kep\ mission through quarter 6 (from which
  this sample is drawn) is roughly delineated with the thin dashed
  lines.} 
\label{planet_stats}
\end{figure*}

The long-term stability evident in our simulation of
\S\,\ref{architecture} allows us to view the present-day architecture
of the \ktt\ planets as representing the state of the system at the 
end of its formation epoch, $\sim 10-100$\,Myr after the formation of
the star. We now investigate what can be learned about the formation
of the \ktt\ planets by looking backward from their end state.

\subsubsection{The \ktt\ Protoplanetary Nebula}\label{mmpn}
Planets form from flattened disks of dust and gas circulating around
protostars, and the masses and sizes of these protoplanetary disks as
well as their lifetimes and accretion rates are constrained from 
observations \citep[see][]{wil11}. The surface
density profiles are typically assumed to follow a power law form
$\Sigma_d \propto a^{-\gamma}$. (Note: Here and throughout
  we use $a$ to denote both the orbital radius of circumstellar
  material and the semimajor axis of the planets. The \kep\ planets
  typically have low eccentricities \citep{wu12} minimizing
  ambiguity.)

The value of $\gamma$ is often taken to be $3/2$ based on estimates of
the “minimum mass solar nebula” \citep[MMSN;][]{wei77,hay81}. The MMSN
is constructed by smoothing the planets of the Solar System over their 
respective domains and correcting for Solar abundance to recreate a
minimal surface density profile from 
which the Solar System could have formed. Modern observations
constrain the surface density profile in the outer disk ($\gtrsim
40$\,AU) to be on average shallower than this,
$\gamma \approx
0.4$--1.0 \citep{and07a,ise09,and09}, while the form of the surface
density profile in the inner disk remains largely unconstrained.  

We can construct a ``minimum mass protoplanetary nebula'' for \ktt\ in
a similar fashion. Since we do not have mass estimates for all the 
planets in the \ktt\ system we use the mass-radius relationship for
Solar System planets $M \propto R^{2.06}$ \citep{lis11}, consistent
with the nominal masses from \S\,\ref{TTV} to within 50\%. 
Assuming a gas-to-dust ratio of 100 and a rocky core
mass fraction of 50\% implies that the protoplanetary nebula of
\ktt\ contained $\approx 3\,M_{Jup}$ of gas between 0.013 and
0.13\,AU of the host star. 

This translates to a very high surface density, $\approx
5\times10^5$\,\gcms, in the inner regions of the disk that is
incongruous with the masses and surface density profiles in
the outer regions of disks as measured with 
(sub)-millimeter interferometry \citep{and07a,ise09,and09}. Indeed, if  
this surface density is used to scale the median best fit values of
the disk radius and surface density profile index for stars less than
1\,\msun\ in the \cite{and09} sample---$R_d = 126$\,AU and $\gamma =
0.9$---the total mass of the \ktt\ protoplanetary nebula would be
orders of magnitude greater than the star! This implies one of three
possibilities: the disk surface density deviated significantly from a
single powerlaw in the inner regions with a large pile-up of material
near the disk inner radius \citep[see, {\eg},][]{chi12}; the material
that formed the planets came 
from elsewhere in the disk; or the planets themselves formed elsewhere
and were transported to their present-day locations.

\subsubsection{Oligarchic Growth and the Isolation Mass}
The gravitational influence of the host star poses another barrier to
super-Earth-sized planets forming within a tenth of an AU. The amount
of material accessible to a growing planetary embryo during the
oligarchic phase of growth is estimated by the ``isolation mass''
\citep{lis87}. This quantity is derived by calculating the amount of
mass available to an oligarch within a disk of planetesimals. We
choose the radius of gravitational influence to be $4$ Hill radii
based on the approximate spacing of isolated oligarchs in numerical
simulations \citep{kok98} to obtain
\begin{equation}
M_{iso} = \frac{\left[16\pi a^2 \Sigma(a)\right]^{3/2}}{(3M_\star)^{1/2}}
\label{miso}
\end{equation}
where $a$ is the radial distance from the star, $M_\star$ is the mass
of the star, and $\Sigma(a)$ is disk surface density profile. To
calculate values for $M_{iso}(a)$, we again use the median surface
density profile of the M stars in the \cite{and09} sample, 
the median values of the M stars in the \cite{ise09} sample with
$\gamma = 0.5$ and $R_d = 260$\,AU, and also the \cite{eis12} model
with a median disk radius of $R_d = 250$\,AU and a surface density
profile with fixed $\gamma = 1.37$ based on theoretical arguments. 
 
Figure~\ref{planet_stats} shows that the isolation mass computed for
these surface density models are $\gtrsim 3$ orders of magnitude too
small to account for the mass in the \ktt\ planets. 
While the planetary embryos can undergo an additional phase of growth
through subsequent mergers that increase their mass by an order of
magnitude \citep[{\eg},][]{cha98}, it would be
difficult to achieve an additional factor of $\sim 100$ 
growth during this phase \citep[see also][]{lis07}. The total amount
of growth in this late phase growth is limited to the total amount of
mass present in the disk at the outset, which is addressed in the
previous section.

The migration of planetesimals from outside this region also offers a
possibility for late time growth of the \ktt\ planets
\citep{han12}. However the gravitational stirring of the planetesimals
can cause radial displacements of the planetary embryos away from
period commensurability. The magnitude of this effect has recently
been used to place limits on the amount of mass contained in
orbit-crossing planetesimals at the end stages of planet formation for
closely spaced, low inclination multi-planet \kep\ systems
\citep{moo12}.
These limits of $\sim 1$\,\mearth\ are too low to contribute
significantly to the discrepancy between the isolation mass and the
final mass of the \ktt\ planets.   

This corroborates the above conclusions that the \ktt\ protoplanetary
disk either contained a disproportionately large amount of mass within
0.1\,AU, or that the formation of the \ktt\ planets followed a
more complex formation and evolution history.

\subsubsection{The Dust Sublimation Radius}
Planets form from dust in protoplanetary disks. But at small orbital
radii the stellar irradiation raises the dust temperature beyond
the sublimation point, clearing these regions of the raw materials for 
planet formation. Low-mass stars such as \ktt\ begin pre-main
sequence evolution with luminosities much larger than their main
sequence values and take $>10$\,Myr to contract to the
point when hydrogen burning begins. The sublimation radius at
these early epochs is thus further from the star than one would
conclude from the present-day luminosity of the star.

\begin{figure}
\centering
\includegraphics[angle=0,width=3.4in]{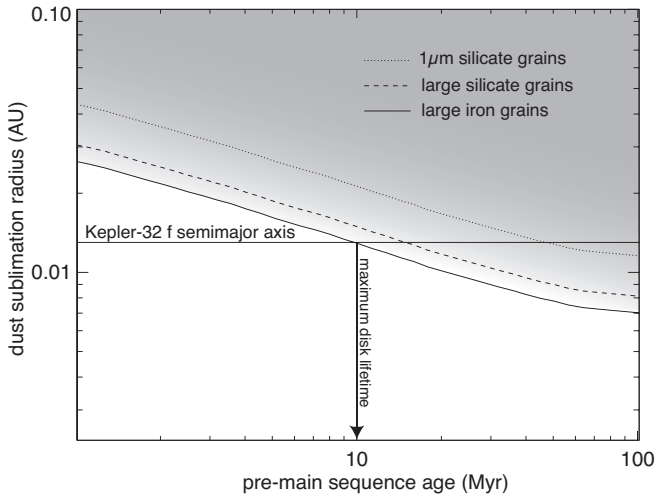}
\caption{Dust sublimation radius of \ktt\ as a function of pre-main
  sequence age according to the models of \cite{bar98}. The solid
  curve is for large iron grains ($T_{sub} = 1908$\,K and $\epsilon
  = 1$) the dashed curve is for large silicate (olivine) grains
  ($T_{sub} = 1774$\,K and $\epsilon = 1$), and the dotted curve is
  for 1\,$\mu$m silicate (olivine) grains within a lower density
  environment ($T_{sub} = 1570$\,K and $\epsilon = 0.58$). All curves
  cross the present day semimajor axis of \ktt\,f after the typical
  lifetime of a protoplanetary disk. }
\label{subfig}
\end{figure}

The dust sublimation radius is given as
\begin{equation}
R_{sub} = 0.034\left(\frac{1500}{T_{sub}}\right)^2
\sqrt{\frac{L_\star}{L_\odot}\left(2+\frac{1}{\epsilon}\right)}
\label{rsub}
\end{equation}
\citep{ise09} where $T_{sub}$ is the sublimation temperature of the dust,
$L_\star$ is the stellar luminosity, and $\epsilon$ is a measure of
the cooling efficiency of the grains
with larger grains having higher efficiencies leading to smaller
sublimation radii \citep{ise06}. Figure~\ref{subfig} shows the
evolution of $R_{sub}$ as 
a function of pre-main sequence age of a 0.5\,\msun\ star based on the
stellar evolution models of \cite{bar98}.

Even if it is conservatively assumed that the dust is in large grains
($\epsilon = 1$) and composed of pure iron with a sublimation
temperature $T_{sub} =1908$\,K \citep{pol94}, the sublimation radius
does not move inward of \ktt\,f's orbit until after 9.8\,Myr of
pre-main sequence evolution. For smaller grains made of silicates and
for less dense 
environments the dust sublimation radius lies beyond the semimajor axis
of \ktt\,f until after $\sim 50$\,Myr, and beyond the present day
orbit of \ktt\,e for $\sim 3$\,Myr of pre-main sequence
evolution. This poses a serious challenge to {\em in situ} formation
particularly since the dusty disks from which planets form only last
$\lesssim 10$\,Myr before they are drained, disrupted, or evaporated away
\citep{wil11}.


\subsubsection{{\em In Situ} Formation vs. Formation then Migration} 
\label{insitu}
The above investigation into the structure of the \ktt\ planetary
system as well as the history and evolution of the host star present 
meaningful constraints on its formation. 
To form \ktt\,f and e where we view them today requires that they
formed from planetesimals scattered inward from the wider orbit
planets. This kind of formation scenario would be different than the
theory described by \cite{han12} where scattered planetesimals accrete
onto pre-existing 
oligarchs from an early phase of planet formation. Rather \ktt\,f and
e would have had to have formed from scattered planetesimals in a
region largely 
devoid of solid material. It is not clear how feasible this would be,
and requires an investigation beyond the scope of this work. An
additional challenge for {\it in situ} formation of the \ktt\ planets
is to explain the near period commensurabilities of \ktt\,e, b and c
with {\it in situ} formation that we have shown to be statistically
significant beyond pure chance. 

On the other hand, a scenario in which the \ktt\ planets formed
further out in the disk and then migrated to their present locations
offers a natural explanation for 1) the large amount of planetary mass
within 0.1\,AU of this 0.54\,\msun\ star, 2) the position of \ktt\,f
within the dust sublimation radius of the star for $\sim 10$\,Myr of
pre-main sequence evolution, and 3) the near period commensurabilities
between three of the planets. The high volatile content inferred for
\ktt\,b and c is also indirect evidence that these planets 
may have acquired volatiles from beyond the snow line at $\sim 1$\,AU
\citep{ken08b}. The difficulties in applying this concept lie in
understanding the mechanisms of migration through a disk of gas and
planetesimals. While a full, rigorous analysis of these possibilities
is also beyond the scope of this article, we look to some order of
magnitude calculations of different migration mechanisms for insight.

\subsubsection{Migration}
The compact and coplanar architecture of the \ktt\ system favors
migration through a disk rather than through planet-planet
interactions. Interactions with either a gaseous 
disk or a disk of planetesimals can cause radial migration of
planetary orbits \citep[see][for a review]{kle12}. Planets massive
enough to open a gap in the protoplanetary disk are coupled to the
viscous evolution of the disk while less massive planets migrate more
quickly due to larger disk torques.

Using a generalized criterion for gap formation in a viscous disk
\citep[][Eq. 15]{cri06} we find that none of the \ktt\ planets would
have opened a gap in their protoplanetary disk assuming a constant
disk flaring term $a/H =0.05$ where $H$ is the scale height, a
viscosity parameter $\alpha = 10^{-3}$ and using the nominal masses
for \ktt\,b and c (\S\,\ref{TTV}) and $M_p\propto R_p^{2.06}$
otherwise. We therefore restrict our discussion of migration through a
gaseous disk to the ``type I'' mechanism \citep{gol79,lin79} 


Assuming a protoplanetary disk with  with a powerlaw surface density
profile, $\Sigma = \Sigma_0a^{-\gamma}$, and a midplane temperature
profile, $T \propto a^{-\beta_T}$, we estimate the migration rate 
through a gaseous disk following \cite{kle12} 
\begin{equation}
\frac{da}{dt} = \Gamma_{\rm tot}\frac{2}{m_p}\sqrt{\frac{a_p}{GM_\star}}
\label{typeI}
\end{equation}
where $p$ subscripts refer to planet quantities, and 
$\Gamma_{\rm tot}$ is the total torque on the planet from Linblad,
corotation, and horseshoe torques
\begin{equation}
\Gamma_{\rm tot} = \left[-1.36 + 0.62\gamma+0.43\beta_T +
  1.36(3/2-\gamma)\right]\Gamma_0 
\label{torque}
\end{equation}
\citep[also see][]{mas06b}, and
\begin{equation}
\Gamma_0 =
\left(\frac{m_p}{M_\star}\right)^2\left(\frac{H}{a_p}\right)^{-2}
\Sigma_pa_p^4\Omega_p^2 .
\label{Gamma0}
\end{equation}
As a baseline model we use a disk surface density profile  
with $\gamma = 1$, which extends from the present day location of
\ktt\,f out to 200\,AU, contains a total of 10\% of the stellar mass
and has a midplane temperature profile with $\beta_T = 3/4$. This disk
model yields type I migration rates $da/dt \sim -0.07, -0.3,
-0.7, -0.9$ and $-1$\,AU\,($10^4$\,yr)$^{-1}$ for \ktt\,f through
\ktt\,c, respectively with the migration rate scaling as  $da/dt
\propto m_p\Sigma_0a_p^{3/2-\gamma}$. For these calculated rates, the
timescale for these planets to have migrated from beyond the snow line
at $\sim 1$\,AU \citep{ken08b} to their present day locations is
short, $\tau_{mig} \sim 10^4$\,years, in comparison to typical disk
lifetimes.

The increasing mass of the \ktt\ planets as a function of semimajor
axis naturally produces convergent migration for type I torques in a
smooth disk, a necessary but not sufficient condition for resonant
capture. To estimate the probability for resonant capture we perform
an order of magnitude calculation comparing the resonance crossing
times to the libration time for the resonances. The resonance widths
and libration times are estimated using the pendulum model for an
interior resonance \citep[][\S\,8.6-7]{mur00}. For the 2:3 resonance of
\ktt\,b and c, the libration period is $T_{lib} \sim 10$\,years and
and the width is estimated to be $\Delta a_{res} \sim 0.009$\,AU using
the mean eccentricity 
of \ktt\,b from the simulations of \S\,\ref{architecture}. With a
convergence rate of 0.2\,AU\,($10^4$\,yr)$^{-1}$ we find that
the resonance crossing time is roughly 40 libration periods. Following
the same procedure for the 1:2 resonance of \ktt\,e and b, we find
that the resonance crossing is of order 10 libration periods.

Shallower surface density profiles produce slower convergence rates,
increasing the probability of resonant capture while steeper profiles
produce faster convergence, shortening resonance crossing times and
decreasing the probability for capture. For $\gamma = 1/2$ we find
that the resonance crossing times are on the order of $10^3$ libration
periods for both resonances and for $\gamma = 3/2$ these times are
shortened to $\lesssim 1$. The resonance crossing times in units of
libration periods is a weakly increasing function of semimajor axis
for our simple model, going as $a_p^{1/4}$. 

Therefore we find that the \ktt\ planets could have been captured into
resonance while migrating through a gaseous disk if the surface
density profile had $\gamma \lesssim 1$, consistent with
modern observations of protoplanetary disks \citep{and09}. The
probability for capture is a weak function of semimajor axis, and it
is thus also feasible for the \ktt\ planets to have been caught in
resonance further out in the disk and have moved inward in lockstep
\citep[{\eg}][]{cre08}. How the planets may have stopped their
migration before falling onto the star remains an outstanding
question in this scenario. 

It is also possible that the \ktt\ planets migrated through a
planetesimal disk \citep{lev07} rather than the gas dominated disk
required for type I migration. We calculate the order of magnitude
rates for planetesimal migration using the equations of \cite{brom11}
assuming a gas to dust ratio of 100 and that all solids are in the
form of planetesimals. We find migration rates of $da/dt \sim -0.44,
-1.1, -1.7, -2.1, -3.2$\,AU\,($10^8$\,yr)$^{-1}$, roughly 4 orders of
magnitude slower than type I. This slow migration rate would favor
resonance capture and may circumvent the need for an efficient
stopping mechanism. However, it may be difficult to migrate these
planets in lock step from much further out in the disk as the inner
planets could stir the planetesimal disk reducing the migration
efficiency at that location for the next further out planet
\citep{kir09,brom11}. Also if \ktt\,b and c were migrating at such
slow rates, it is likely that they would have ended up in a 1:2
resonance rather than the more compact 2:3 resonance. These details
may limit the distances over which planetesimal migration could have
acted to produce the observed architecture of \ktt.

From these analyses, migration through a gaseous disk produces a more
favorable situation for transporting the \ktt\ planets from far enough
out in the disk to reconcile the difficulties with {\it in situ}
formation outlined above. However, further investigation into the
migration of multi-planet systems through both gas and
planetesimal disks is needed to draw more definitive conclusions.


\section{Discussion}\label{discussion}
The analyses of the preceding sections provide evidence that the 
\ktt\ planets formed further out in their protoplanetary disk from
where we see them today and migrated convergently into their present
locations. The high inferred volatile content of \ktt\,b and c, and
the order of magnitude calculations for the migration rates of the
\ktt\ planets suggest that these planets have formed beyond the snow
line in the 
presence of gas. If true, the \ktt\ planets would have necessarily
formed within $\sim 10$\,Myr, the timescale over which gas survives in
protoplanetary disks \citep{wil11}. This formation history stands in
contrast to the formation of the terrestrial planets in the Solar
System, which are commonly thought to have formed {\em in situ} in a
gas-free environment on a 100\,Myr timescale \citep{wet90}. The
possibility to constrain the timescale for the formation of the
\ktt\ planets motivates further investigation into migration
of multi-planet systems.  

Our conclusions about the formation of the \ktt\ planets rely on a
detailed characterization of a single system not possible for the full
ensemble of {\kep}'s M dwarf planets. However, both
the \ktt\ star and its planets are representative of this full
ensemble giving important clues regarding the formation of full sample
of \kep\ M dwarf planet candidates.

\subsection{The Ensemble of \kep\ M Dwarf Planets} \label{ensemble}
Our sample of M dwarfs was drawn from the catalog of \kep\ Objects of
Interest given by \cite{bat12} using a color magnitude cut of $K_P >
14$ and $K_p - J > 2$ \citep{man12}. This results in 5499 stars from
the KIC catalog observed with \kep, however there are only 4682 with
finite, non-zero photometric precision values for at least one quarter
between Q0 and Q6. 
Since the \kep\ sample is magnitude limited, the sample is skewed
toward the most massive M dwarfs, peaking around 0.5\,\msun (see
Figure~\ref{star_dists}). Two giant planet candidates 
are present in this sample. One is a {\em bona fide} giant planet
confirmed with radial velocity data  
\citep[KOI-254;][]{joh12}. The other, KOI-1902, we found to be a
false positive based on its transit profile. We scrutinized the light
curves of several of the other apparent outliers by hand. We reject
two KOIs in the list, KOI-531 and KOI-1152, on account of deep
secondary transits, and two others, KOI-256 (Muirhead et al. in
prep.), and KOI-1459 based on 
their transit profiles. The final distribution of M dwarf  
planet candidates are shown as gray circles in
Figure~\ref{planet_stats}. These remaining candidates are expected to
have a high fidelity allowing us to treat them as a
sound, statistical ensemble. 

The distribution of M dwarf planet candidates in
Figure~\ref{planet_stats} contains 100 planets in 66 systems. There
are 48 single planet systems, 7 double planet systems, 7 three planet
systems, 3 four planet systems, and 1 five planet
system---\ktt. Therefore 18 systems or 27\% of the M dwarf KOIs are
multi-transit systems. The giant, KOI-254 is a single candidate system
that constitutes 1\% of all planets in the \kep\ M dwarf sample and about
2\% of all planet hosting M dwarfs. This planet is anomalous and we
do not consider it as part of the ensemble hereafter. 

The main distribution of M dwarf planets appears to follow a trend of
increasing mass (radius) as a function of semimajor axis. Without
accounting for biases, we derive the relationship, $M \propto
a^{0.6}$. However, for a given signal to noise threshold the lower
envelope of planet candidates is expected to follow 
\begin{equation}
R_{lim}= ({\rm
  SNR}\,\sigma_p)^{1/2}R_\star^{3/4}a^{1/4}\left(\frac{\pi}{n}\right)^{1/4}
\label{sn}
\end{equation}
where SNR is the signal to noise threshold, $\sigma_p$ is the
photometric precision, $R_\star$ is the stellar radius and $n$ is the
number of transits measured. For our adopted mass-radius relationship
this translates to $M_{lim} \propto a^{0.5}$. The  
detection limit for an observational baseline of 6 quarters and a SNR
threshold of about 5 for a typical M1V star is shown in
Figure~\ref{planet_stats} as the dashed line. This suggests that the
apparent trend at the lower envelope of the distribution is 
due to observational biases. Also shown in Figure~\ref{planet_stats}
is the semimajor axis corresponding to \kep\,\!'s observing time
baseline over the first 6 quarters of data over which the planet
candidates were selected \citep{bat12}.

\begin{figure}
\centering

\includegraphics[angle=0,width=3.4in]{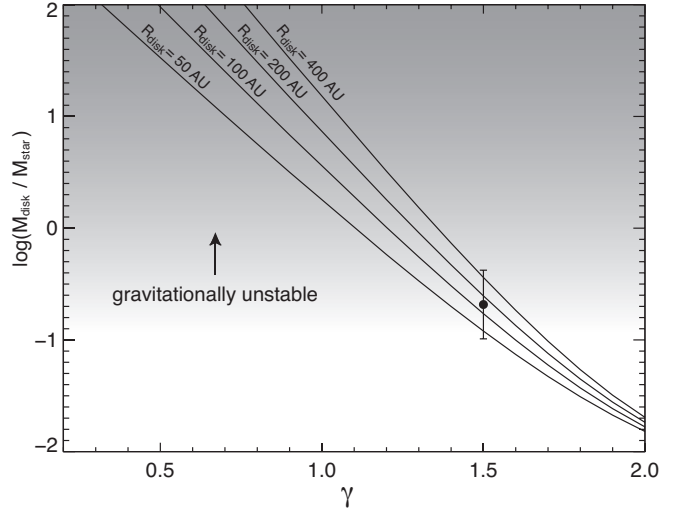}
\caption{The inferred disk mass in units of the stellar mass for all
  the M dwarf multi-transit systems in the \kep\ sample as a function
  of assumed surface density profile index $\gamma$, and disk radius,
  $R_{\rm disk}$. The data point and error bars show the mean and
  standard deviation of one such calculation for $\gamma = 1.5$ and
  $R_{\rm disk} = 150$\,AU to give a sense of the distribution of disk
  masses derived in each calculation. Disks with masses $\gtrsim
  0.2\,M_\star$ are expected to be gravitationally unstable.}
\label{MMPD}
\end{figure}

Although we cannot perform as detailed analysis on the \kep\ M dwarf
planet ensemble as we have for \ktt, we note that positions and
inferred masses of the planet candidates as a whole also imply very
large disk masses within 0.1\,AU \citep[also see][]{chi12}.
Figure~\ref{MMPD} shows the reconstructed mean protoplanetary disk
masses for the 18 multi-transit M dwarf KOIs as a function of 
$\gamma$, the 1\,$\sigma$ spread in the derived masses for any given
value of $\gamma$ is approximately 0.3 dex (shown as data point and
error bar). Even for $\gamma = 3/2$, an appreciable fraction of the
disks would be gravitationally unstable and therefore unlikely to
produce the observed planets where we see them.

We also analyze the present-day locations of the \kep\ M dwarf planets
in terms of the sublimation radius of their host star at 10\,Myr of
pre-main sequence evolution. We find that between 5 and 14\% of the
total number of planet candidates fall within the dust sublimation
radius of their host stars for large iron grains and $1\,\mu$m
silicate grains, respectively, and could not have formed in place.

\subsection{\ktt\ as a Representative of the \kep\ M Dwarfs}
\begin{figure}
\centering
\includegraphics[angle=0,width=3.4in]{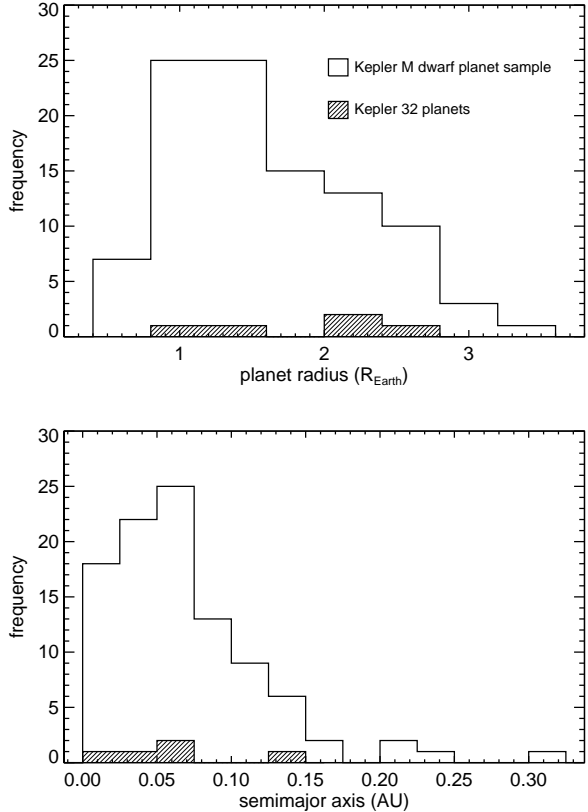}
\caption{The distributions of \kep\ M dwarf planet candidates as a
function of planet radius and semimajor axis (blank histogram; not
including KOI-254, 256, 531, 1152, 1459, or 1902) in comparison to the
\ktt\ planets (hashed histogram).} 
\label{planet_dists}
\end{figure}

The \ktt\ planets span the main distribution of the \kep\ M dwarf
planets seen in Figure~\ref{planet_stats}
and we plot in Figure~\ref{planet_dists} the locations of these
planets in relation to the full distribution as a function of planet
radius and semimajor axis. In both parameter spaces the \ktt\
planets fall well within 
the main distribution of planet candidates.

To further explore how
representative the \ktt\ planetary system is of the full sample of
\kep\ M dwarf planet candidates, we create an ensemble of planetary
systems with the \ktt\ system specifications oriented randomly on the
sky with mutual inclinations drawn randomly from a Rayleigh
distribution \citep{lis11}. We find the transit multiplicity fractions
are best reproduced with a spread in mutual inclinations of $1.2^\circ
\pm 0.2^\circ$, consistent with values for the entire \kep\ planet  
candidate ensemble \citep[$1.0^\circ$--$2.3^\circ$;][]{fab12b}. The
discrepancies between the real and   
simulated distributions of transiting systems are largest for the
single transit systems (55\% simulated vs. 73\%
real). This would be  
expected for a situation where the typical 
planetary system contains less than 5 planets or is less
compact. However, we note that the observed transit multiplicity of
\kep\ M dwarf systems can be recreated remarkably well assuming the
case that all planetary systems are exact clones of \ktt.

\begin{figure}
\centering
\includegraphics[angle=0,width=3.4in]{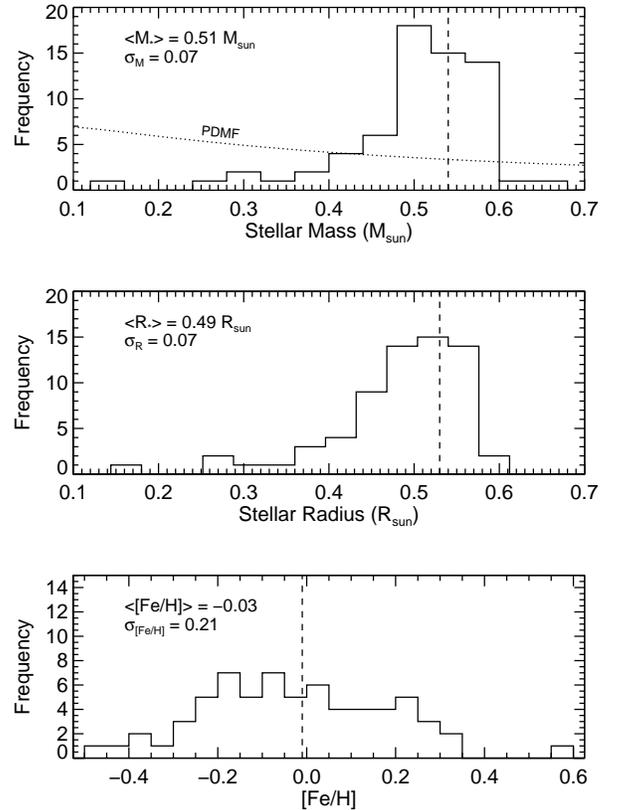}
\caption{The distribution of stellar masses and radii
  for our sample of M dwarf planet host stars are sharply peaked
  around 0.53\,\msun\ and 0.51\,\rsun, respectively. The present day
  mass function of single field stars in the Galaxy \citep{cha03}
  normalized to the number of stars in our sample is shown as the
  dotted line in the top panel for reference. Metallicities for our
  sample fall mostly between $-0.4$ 
  and $0.4$. \ktt\ values denoted by the vertical dashed line all fall
  near the center of these distributions. } 
\label{star_dists}
\end{figure}

The distribution of stellar masses, radii, and metallicities for the
\kep\ M dwarf sample are shown in Figure~\ref{star_dists}. A narrow
range of stellar mass and radius is seen peaked around 0.5 in solar
units due to the magnitude limit of the sample and is shown in
contrast to the present day mass function of single stars
\citep[PDMF;][]{cha03} in the top panel. \ktt\ is seen to be
representative of the 
full sample in all quantities.

\subsection{Planet Occurrence}
The planet occurrence rate for M dwarf planets has been calculated to
be about 0.3 for planets with radii $> 2$\,\rearth\ and periods less
than 50 days \citep{man12}. However, only 28 planets of the 100 total
in our \kep\ M dwarf ensemble satisfy these criteria. Therefore the
total planet occurrence rate for short period planets around M dwarfs
is much higher than this number.
The detection of planet signals within the \kep\ M dwarf
sample is not uniform causing a systematic uncertainty to planet
occurrence estimations. Since we have no way to correct for this
currently, we ignore this effect which will result in a lower limit of
the occurrence rate. 

The number of
planets per star in our sample is estimated as
\begin{equation}
f = \sum_i^{n_p} \frac{1}{\sum_j^{n_{\star,i}}p_{i,j}}
\label{poccur}
\end{equation}
where $j$ is the index of the sum over all stars around which planet
$i$ could be detected, $n_{\star,i}$ \citep[see, {\eg},][]{how12}, and 
\begin{equation}
p_{i,j} = \frac{R_{\star,j}}{a_i}
\label{prob}
\end{equation} 
is the geometric probability of detecting planet $i$ around star
$j$ if eccentricity is negligible and $R_p/R_\star \ll 1$. Since the KIC
stellar radii values for stars with  
$T_{\rm eff} \lesssim 4500$\,K are unreliable, we randomly sample the
distribution of stellar radii in Figure~\ref{star_dists} estimated
from near infrared spectra for this calculation. Evaluating
Equation~\ref{poccur} over our sample 
yields $f = 1.0 \pm 0.1$ planets per star where we use a binomial
error estimate. 

Though we do not perform statistical validations for
all the \kep\ M dwarf planets, our sample is expected to have a higher
fidelity than the total ensemble of \kep\ transit signals since we
culled our sample by hand. The planet occurrence derived from our
sample assuming 90\% fidelity, $f_{90}$, is calculated by evaluating
equation~\ref{poccur} repeatedly for 90 randomly drawn planets of the
100 total. We find $f_{90} = 0.9 \pm 0.1$ where the stated error is
the standard deviation of the distribution of 1000 realizations. As a 
futher check, we limit our sample to the 28 planets with $R_p >
2$\,\rearth\ and $P < 50$\,d to obtain 
$f = 0.26 \pm 0.05$ at the lower end of the estimations of
\cite{man12}. This may be further evidence that our estimations are
conservative. 

Counting only one planet per system in equation~\ref{poccur} gives the
occurrence of stars with at least one planet, $0.51\pm0.07$
(not including KOI-254), which together with our estimate of the total
occurrence gives the average number of planets per system as $\sim
2$. Stars with masses characteristic of the \kep\ M dwarf sample are a
factor of $\sim 1.8$ times more common than 
stars of 1\,\msun\ \citep{cha03}, and the planets of the \ktt\ system
are representative of the planets that form around these
stars. Therefore the insights gleaned from the \ktt\ system both show
us where to look for additional planets in the Solar Neighborhood, and
provide a template for understanding the 
formation of the ubiquitous compact planetary systems throughout the
Galaxy. 

\subsection{Summary and Future Directions}
We present a detailed analysis of the \ktt\ planetary system which
offers the rare circumstance of 5 transit signals. While two of the
planets have been previously validated through evidence of their
mutual gravitational interactions, we validate the remaining three
transit signals probabilistically using observations from the
W. M. Keck Observatory as 
constraints.  This validation makes \ktt\ the richest system of
transiting planets known around an M dwarf. 

\ktt\ has a markedly compact architecture. All five planets orbit
within one third of Mercury's distance from the Sun, with the closest
planet orbiting only 4.3 stellar radii from the \ktt\ photosphere. The
three middle planets lie close to a 1:2:3 period commensurability that
is unlikely to be the result of chance. 

Our refined stellar parameters improve the derived planetary
characteristics, and aid in reconstructing this system's formation
history. Several pieces of evidence from our analyses indicate that
the \ktt\ planets did not form where we see them today:
\begin{itemize}
\item the dust sublimation radius of \ktt\ lying outside the
  present day semimajor axis of \ktt\,f for longer
  than a typical protostellar disk lifetime
\item the extremely high surface densities inferred by assuming {\em
  in situ} formation
\item the limited range of gravitational influence for
planetary embryos located so close to their host star
\item the unlikely arrangement of three planets in the system near a
  1:2:3 period commensurability 
\item the high volatile content of \ktt\,b and~c.
\end{itemize}

This conclusion necessitates planet migration through a disk, and our
order of magnitude calculations for the migration rates of the
\ktt\ planets embedded in a typical protostellar disk suggest the
presence of gas. If true, this would limit the formation time of the
\ktt\ planets to $\lesssim 10$\,Myr---the known timescales over which
gaseous disks survive. 

\ktt\ is found to be representative of the full sample of 66 \kep\ M
dwarf host stars, and the \ktt\ planets span the mid-line of the
distribution of 100 \kep\ M dwarf planet candidates in
radius-semimajor axis space. Although we are
unable to treat each system in this ensemble with the same care as for
\ktt, we show that similar analyses applied to the ensemble give
consistent results to those derived for \ktt. Thus the formation
scenario deduced from \ktt\ offers a plausible blueprint for the
formation of the full sample of \kep\ M dwarf planets. 

We select out 4682 M dwarfs from the Kepler Input Catalog that have
been observed with \kep\ and use their
observational parameters to derive the planet occurrence rate of
\kep\ M dwarf planet candidates. We confirm that within the
completeness limits of the first 6 quarters of \kep\ data, the M dwarf
planet candidates have an occurrence rate about 3 times that of
solar-type stars, while the occurrence rate of all candidates around M 
dwarfs is $1.0 \pm 0.1$. We expect the fidelity of our culled sample
to be above 90\%. Thus the compact systems of planets around
the \kep\ M dwarf sample are a major population of planets throughout
the Galaxy amplifying the significance of the insights gleaned from
\ktt. 

At the time of this writing, there are only 37 planets confirmed to
exist around 24 M dwarfs in the Galaxy
\citep[exoplanets.org,][]{wri11}. The \kep\ space telescope  
has revealed 100 planet candidates around 66 M dwarfs from which we
draw our statistics. It would be of great benefit to the study of M
dwarf planet formation to explore and validate a larger statistical
sample of M dwarfs such that comparisons can be drawn against the
detailed analyses of solar-type stars \citep[{\eg},][]{you11}. The
continued monitoring of the current sample is also important as this
will reveal trends in planet occurrence as a 
function of orbital period for the smallest planets. 

Mass measurements of the growing numbers of confirmed M dwarf planets
will also play an important role in interpreting their origins. This
can be achieved with nearby M dwarfs using precision radial velocity
measurements, or by using alternative techniques such the amplitude of
transit timing variations. 

Lastly, direct imaging of the inner few AU of nearby protostellar
disks will be possible in multiple wavebands with the Atacama Large
Telescope Array (ALMA). The modeling of this emission may be the most
direct way to constrain where in the protoplanetary disk compact
planetary systems form. 

\acknowledgements
This work has benefited from the feedback and suggestions of many
people including but not limited to Andrew Youdin, Leslie Rogers,
Peter Goldreich, Hilke Schlichting, and Nairn Baliber. We also thank
the anonymous referee for a thorough review. D.C.F. acknowledges
support for this work was provided by NASA through Hubble Fellowship
grant HF-51272.01-A, awarded by the Space Telescope Science Institute,
which is operated by the Association of Universities for Research in
Astronomy, Inc., for NASA, under contract NAS 5-26555.

Some of the data presented herein were obtained at the W.M. Keck
Observatory, which is operated as a scientific partnership among the
California Institute of Technology, the University of California and
the National Aeronautics and Space Administration. The Observatory was
made possible by the generous financial support of the W.M. Keck
Foundation. 

This paper includes data collected by the Kepler mission. Funding for
the Kepler mission is provided by the NASA Science Mission
directorate. 

The Robo-AO system is supported by collaborating partner institutions,
the California Institute of Technology and the Inter-University Centre 
for Astronomy and Astrophysics, by the National Science Foundation
under Grant Nos. AST-0906060 and AST-0960343, by a grant from the
Mt.Cuba Astronomical Foundation and by a gift from Samuel Oschin.


\end{document}